\documentstyle[epsf]{lamuphys}
\makeatletter
\let\chapter\hid@chapter
\makeatother
\newcommand{\HI}{\mbox{\normalsize H\thinspace\footnotesize I}}

\newcommand{\tfr}{Tully\,--\,Fisher relation}

\newcommand{\kms}{\,km\,s$^{-1}$}
\newcommand{\etal}{{\it et~al.}}
\newcommand{\cf}{{\it cf.\,}}
\newcommand{\eg}{{\it e.g.},\ }         
\begin{document}
\pagenumbering{arabic}
\title{Galaxies behind the deepest extinction layer of the southern Milky Way}

\author{Ren\'ee\,C.\,Kraan-Korteweg\inst{1},
~B\"arbel\,Koribalski~\inst{2}, 
and ~Sebastian\,Juraszek~\inst{3,2}}

\institute{Departamento de Astronomia, 
Universidad de Guanajuato, Mexico \\
kraan@norma.astro.ugto.mx 
\and
Australia Telescope National Facility, CSIRO,
P.O. Box 76, Epping, Australia \\
bkoribal@atnf.csiro.au
\and
School of Physics, Univ. of Sydney, NSW 2006, Australia,
sjurasze@atnf.CSIRO.AU}

\maketitle

\begin{abstract}
About 25\% of the optical extragalactic sky is obscured by the dust and 
stars of our Milky Way. Dynamically important structures might still lie
hidden in this zone. Various approaches are presently being employed to
uncover the galaxy distribution in this Zone of Avoidance (ZOA). Results as 
well as the different limitations and selection effects from these 
multi-wavelengths explorations are being discussed. 
Galaxies within the innermost part of the Milky Way --- typically at a 
foreground obscuration in the blue of  $A_{\rm B} \ga 5^m$ and $|b| \la
\pm5\degr$ --- remain particularly difficult to uncover except for 
H\,{\sc i}-surveys: the Galaxy is fully transparent at the 21cm line 
and H\,{\sc i}-rich galaxies are easy to trace. We will report here on
the first results from the systematic blind H\,{\sc i}-search ($v \leq
12700$ km\,s$^{-1}$) in the southern Zone of Avoidance which is currently
being conducted with the Parkes Multibeam (MB) Receiver. 
\end{abstract}
\section{Introduction}
To understand the dynamics within the local Universe -- the mass
distribution and the local velocity field with its peculiar and 
streaming motions -- a detailed map of the 3-dimensional galaxy 
distribution is highly desirable. However, the dust
extinction and confusion with stars in the disk of our 
Galaxy make this very difficult for $\sim$25\% of the sky, 
and the following questions remain unanswered:

Could a nearby Andromeda-like galaxy
have escaped detection to date, hence change our
understanding of the internal dynamics and mass derivations of the 
Local Group (LG),
and the present density of the Universe from timing arguments (Peebles 1994)?

Is the dipole in the Cosmic Microwave Background
Radiation (direction and amplitude) entirely explained by the gravity
on the LG from the irregular mass/galaxy distribution? 
As the nearest galaxies ($v<300$ \kms) generate 20\% of the total 
dipole moment (Kraan-Korteweg 1989) nearby individual galaxies are 
equally important as massive groups, clusters and voids.

Is the mass overdensity in the Great Attractor (GA) region -- 
postulated from a large-scale systematic flow of galaxies towards  
($\ell,b,v)\sim(320\degr,0\degr,4500$\kms) (Kolatt \etal\ 1995) 
-- in the form of galaxies, hence does light trace mass?

Does the Supergalactic Plane, other superclusters, walls 
and voids connect across the Milky Way and might 
other large-scale structures (LSS) have gone undetected due to 
this 'zone of avoidance'?

\section{Multiwavelengths explorations of the southern ZOA}
Various approaches are presently being employed to uncover the galaxy
distribution in the ZOA: deep optical searches, far-infrared
(FIR), near-infrared (NIR) surveys and blind \HI\ searches. All 
methods produce new results, but all suffer from (different) 
limitations and selection effects. \\
 
\noindent {\bf OPTICAL:}
Nearly the whole southern ZOA has been systematically surveyed for 
highly obscured but still visible galaxies using existing sky surveys
(\cf\ Woudt 1998, for a detailed overview). These surveys achieve
a considerable reduction of the ZOA and have uncovered distinct 
LSS unrelated with the foreground 
extinction. Follow-up redshift observations have revealed a
number of dynamically important structures such as \eg\ the 
nearby overdensity in Puppis (Lahav \etal\ 1993) and the massive
cluster A3627 at the core of the GA (Kraan-Korteweg \etal\ 1996, 
\cf Fig.~1).
Deep optical surveys are not biased with respect to any particular
morphological type. However, for foreground extinctions 
above ${\rm A}_{B} \ga 5^m$
(\HI-column-densities ${\rm N}_{\rm HI}\ga 6\cdot 10^{21}
{\rm cm}^{-2}$), the ZOA remains fully opaque (\cf\ inner contour in Fig.~1).
For the southern Milky Way this corresponds roughly to $|b| \la
\pm5\degr$. \\

\noindent {\bf FIR:}
The IRAS Point-Source Catalog (PSC) has been exploited in the
last decade to identify galaxy candidates behind the ZOA. Using 
different colour selection criteria, galaxy candidates were
followed up by HI radio surveys (\eg\ Lu \etal\ 1990) or 
by inspection of plates (\eg\ Takata \etal\ 1996).
To avoid confusion with Galactic sources, K-band snapshots have
proved very efficient (Saunders \etal\ 1994).
Confirmed IRAS galaxies can be merged with IRAS galaxy samples 
outside the ZOA to produce uniform whole-sky samples for LSS studies. But 
bright spiral and starburst galaxies dominate
these samples.\\

\noindent {\bf NIR:}
The recent near infrared (NIR) surveys, 2MASS (Skrutskie \etal\ 1997) 
and DENIS (Epchtein 1997),
provide complementary data. NIR 
surveys are sensitive to early-type galaxies, are tracers of 
massive groups and clusters missed 
in IRAS and \HI\ surveys, have little confusion with 
Galactic objects and
are less affected by absorption than optical surveys.

In a pilot study, we examined the effeciency of uncovering 
galaxies at high extinctions with DENIS images
(\cf\ Schr\"oder \etal\ 1997 \& Kraan-Korteweg \etal\ 1998):
highly obscured, optically invisible galaxies can 
indeed be traced to lower latitudes ($|b| \ga 1 - 1\fdg5$) than 
deep optical surveys. This is not only of interest in charting
early-type galaxies but also with respect to the combination of
\HI\ data of heavily obscured spiral galaxies detected in blind 
\HI\ surveys (cf. below) with NIR data, and therewith the possibility
to extend the peculiar velocity field into the ZOA via the NIR \tfr.\\


\noindent {\bf {H\thinspace\footnotesize I}:}
In the regions of highest obscuration and infrared confusion 
the Galaxy is fully transparent to the 21-cm line radiation 
of neutral hydrogen. \HI-rich galaxies can readily be found 
at lowest latitudes through detection of their redshifted 
21-cm emission.
Only low-velocity extragalactic sources (blue- and redshifted) 
within the strong Galactic HI emission will
be missed, and -- because of baseline ripple -- galaxies 
close to radio continuum sources.

Until recently, radio receivers were not sensitive and efficient 
enough to attempt large systematic surveys of the ZOA.
In a pilot survey with the late 300-ft telescope of Green Bank,
Kerr \& Henning (1987) surveyed $1.5\%$ of the ZOA and detected 16 
new spiral galaxies.
Since then a systematic shallow search for nearby, massive 
galaxies has been completed in the north (Henning \etal\ 1998), yielding 
five objects including Dwingeloo 1 (Kraan-Korteweg \etal\ 1994). 

\section{The Parkes Multibeam Survey in the southern ZOA}
In March 1997, a systematic blind \HI\ survey began with the 
Multibeam Receiver (13 beams in the focal plane array) at the 
64\,m Parkes telescope in the most opaque region of the southern 
Milky Way ($213\degr \le \ell \le 33\degr$; $|b| \le 5\degr$). 
The ZOA will be surveyed along constant Galactic latitudes in 
23 contiguous fields of length $\Delta\ell=8\degr$. The ultimate 
goal is 25 scans per field where adjacent strips will be offset in
latitude by $\Delta b = 1\farcm5$ for homogeneous sampling. With a total 
observing time of 1500h, we will obtain an effective integration 
time of 25 min/beam with a 3\,$\sigma$ detection limit of 
15\,mJy. Roughly 3000 detections are predicted for the covered 
velocity range of $-1200 \la v \la 12700$\kms\ (Staveley-Smith 1997). 
This allows 
the detection of dwarfs with \HI-masses as low as $10^6$\,M$_{\sun}$ in the 
local neighbourhood, and will be sensitive to normal Sc galaxies well 
beyond the Great Attractor region. As a byproduct, the survey will 
produce a high resolution integrated column density  map of the 
southern Milky Way and a detailed catalog of high velocity clouds
(\cf\ Putnam \etal\ 1998).

\subsection{First Results from the Parkes Multibeam survey}
At the time of this meeting, the whole southern ZOA survey had been 
surveyed twice ($\Delta b = 17\arcmin$, rms $\sim$ 20\,mJy). 
The cubes of the Hanning smoothed data (26\kms\ resolution) 
were inspected visually for 21 of the fields ($220\degr \le \ell \le
4\degr$) and all galaxy candidates with \HI\ fluxes $\ga 100$\,mJy 
were catalogued. 

87 galaxies were uncovered in this way. Four were seen in 
more than one cube. Though galaxies up to 6500\kms\ were identified, 
most of the galaxies (80\%) are quite local (v$<3500$\kms) due to the 
(yet) low sensitivity. In the low-extinction Puppis region (\cf\
Fig.~1), a large fraction of the galaxies and their velocities were 
already known. In the remaining ZOA, about 1/3 have a 
counterpart in NED or the deep optical surveys.

The distribution of the \HI-detected galaxies is shown in the lower
panel of Fig.~1. Here we also display the results by Juraszek \etal\
(in prep.) in the GA region. In this high priority 
area, defined
as $310\degr \le \ell \le 330\degr$, $|b| \le 10\degr$, four scans 
rms ($\sim$ 15\,mJy) were analyzed and 82 galaxies charted.
This area hence probes deeper and finds -- not unexpectedly --
a peak in the velocity distribution between 3000 and 4500\kms\
in the GA direction.
\begin{figure}[t]
\hfil \epsfxsize 10cm \epsfbox [108 172 476 636] {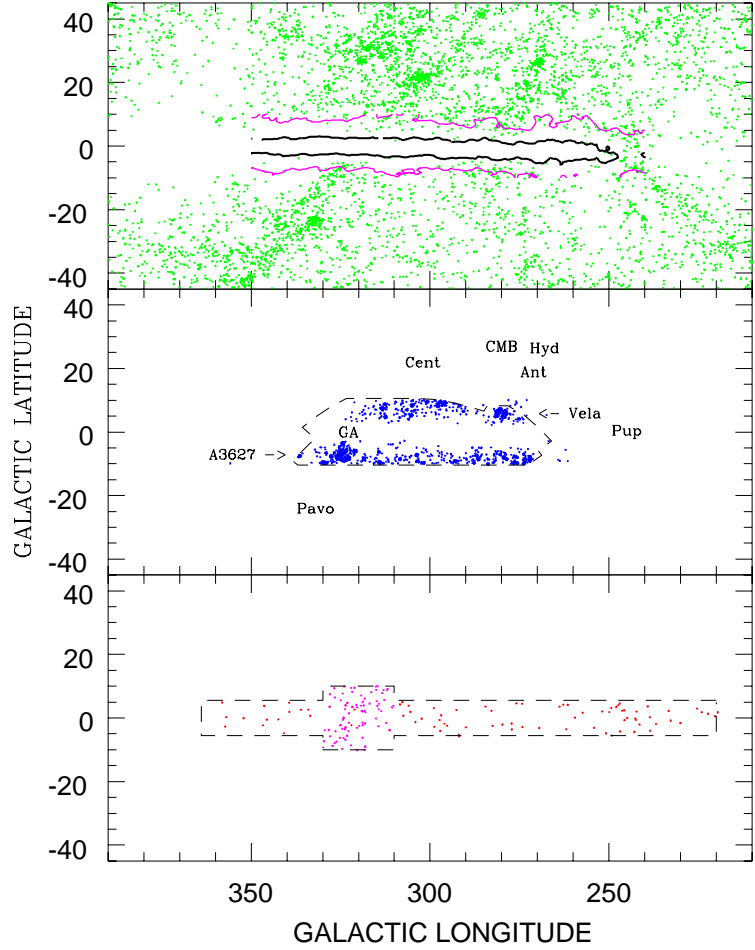} \hfil
\caption[]{Galaxies with v$<$10000 \kms.
Top panel: literature values (LEDA), superimposed
are extinction levels A$_B\sim1\fm5$  and $5^m$; middle panel:
follow-up redshifts (ESO, SAAO and Parkes) from deep optical
ZOA survey with locations of clusters and dynamically important
structures; bottom panel: redshifts from shallow MB-ZOA and
deeper GA survey in \HI\ with the Parkes radio telescope. }
\end{figure}
The top panel of Fig.~1 shows the distribution of all galaxies with
velocities v $\le 10000$\kms\ centered on the southern Milky Way. 
Note the near full lack of galaxy data for extinction levels
A$_B\sim1\fm5$ (outer contour).
The middle panel results from the follow-up observations of the 
optical galaxy search by Kraan-Korteweg and collaborators. Various
new overdensities become apparent at low latitudes. But
the innermost part of our Galaxy remains obscured (A$_B\ga5^m, 
|b|\la5\degr$). Here, the 
blind \HI\ data (\cf\ lower panel) finally can provide the missing link
for LSS studies. 

\begin{figure}[t]
\hfil \epsfxsize 10cm \epsfbox [108 172 476 636] {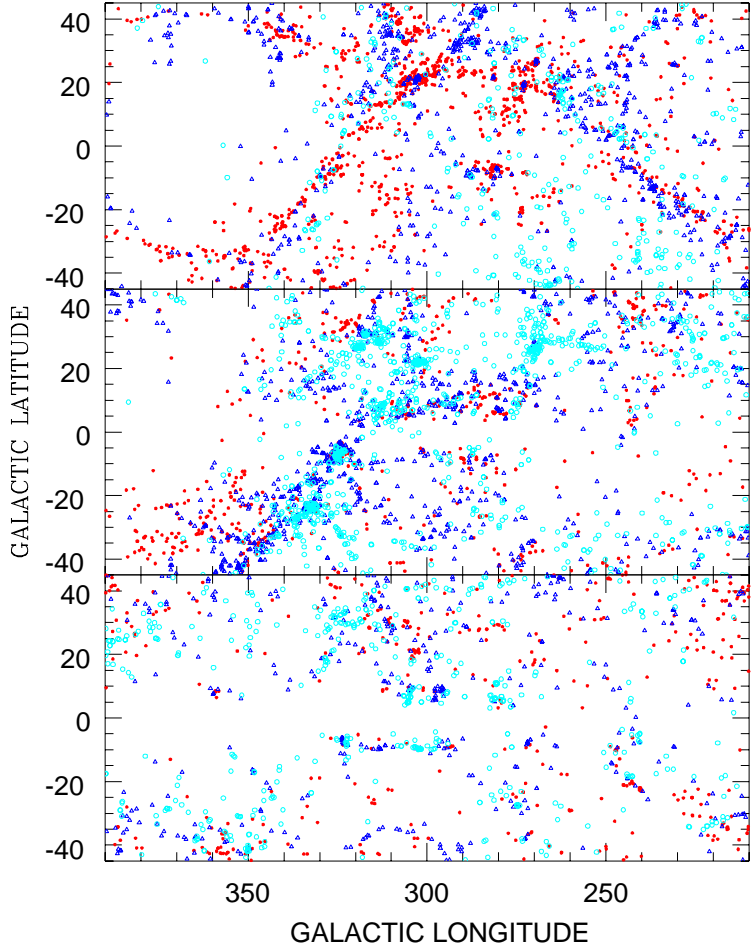} \hfil
\caption[]{Redshift slices from data in Fig.~1: 500$<$v$<$3500
(top), 3500$<$v$<$6500 (middle), 6500$<$v$<$9500 \kms\
(bottom). The open circles mark the nearest $\Delta$v=1000\kms\ slice
in a panel, then triangles, then the filled dots the 2 more distant ones.}
\end{figure}

In Fig.~2, the data of Fig.~1 are combined in redshift slices. The 
achieved sensitivity in the current MB \HI-survey fills in structures
all the way across the ZOA for the upper panel (v $<$ 3500\kms) for the first 
time. Note the continuity of the thin filamentary sine-wave-like 
structure that dominates the whole southern sky, and the prominence 
of the Local Void. 
This feature is very different from the thick, foamy Great Wall-like 
structure, the GA, in the middle panel. 
With the full sensitivity aimed at with the MB-survey, we will be able
to fill in the LSS in the more distant panels of Fig.~2 as well. 

ATCA follow-up observations of three very extended ($20\arcmin$
to $\ga 1\degr$), nearby (v $<$ 1500\kms) sources revealed them to
be interesting galaxies/complexes, with unprecedented 
low \HI\ column densities (\cf\ Staveley-Smith \etal\ 1998).

\section{Conclusions}

The combination of the complementary multiwavelength surveys 
allow a new probing of LSS in the 'former' ZOA. The 
\HI\ surveys are particularly powerful at the lowest latitudes. 
But future merging of ZOA data with catalogs outside the ZOA 
will have to be done with care to obtain 'unbiased' whole-sky surveys.


From the sensitivity attained with the first 2 scans of the 
ZOA MB-survey it can be maintained that no Andromeda or other 
\HI-rich Circinus galaxy is lurking undetected 
behind the extinction layer of the southern Milky Way.\\

{\sl Acknowledgements} --- The help of the HIPASS ZOA team members
R.D. Ekers, A.J. Green, R.F. Haynes, P.A. Henning, R.M. Price, 
E. Sadler, and L. Staveley-Smith is gratefully acknowledged.
%
%

\end{document}